\documentclass[12pt]{article}

\usepackage{epsf}
\usepackage{amsmath}
\allowdisplaybreaks

\begin{document}

\baselineskip=17pt plus 0.2pt minus 0.1pt

\makeatletter
\@addtoreset{equation}{section}
\renewcommand{\theequation}{\thesection.\arabic{equation}}
\renewcommand{\thefootnote}{\fnsymbol{footnote}}

\newcommand{\ap}{\alpha'}
\newcommand{\p}{\partial}
\newcommand{\nn}{\nonumber}
\newcommand{\K}{K_1}
\newcommand{\Me}{M_0}
\newcommand{\Mo}{M_1}
\newcommand{\ve}{\bm{v}_0}
\newcommand{\vo}{\bm{v}_1}
\newcommand{\go}{g_{\rm o}}
\newcommand{\calS}{{\cal S}}
\newcommand{\calQ}{{\cal Q}}
\newcommand{\calEc}{{\cal E}_{\rm c}}
\newcommand{\calNt}{{\cal N}_{\rm t}}
\newcommand{\calQB}{{\cal Q}_{\rm B}}
\newcommand{\calM}{{\cal M}}
\newcommand{\calO}{{\cal O}}
\newcommand{\mt}{m_{\rm t}}
\newcommand{\Phit}{\Phi_{\rm t}}
\newcommand{\phit}{\phi_{\rm t}}
\newcommand{\vac}{\ket{0}}
\newcommand{\Psic}{\Psi_{\rm c}}
\newcommand{\phic}{\phi_{\rm c}}
\newcommand{\calNc}{{\cal N}_{\rm c}}
\newcommand{\resint}[1]{\oint\frac{d #1}{2\pi i}}
\newcommand{\ds}{\displaystyle}
\newcommand{\Pm}{P}
\newcommand{\Qm}{Q}
\newcommand{\Po}{P_{\rm H}}
\newcommand{\Pe}{P_{\rm I}}
\newcommand{\Qo}{Q_{\rm H}}
\newcommand{\Qe}{Q_{\rm I}}
\newcommand{\Lo}{\Lambda_{\rm H}}
\newcommand{\Le}{\Lambda_{\rm I}}
\newcommand{\q}{\bm{u}}
\newcommand{\Kabs}{\sqrt{\K^2}}
\newcommand{\Kabsinv}{\frac{1}{\Kabs}}
\newcommand{\biginv}{{\cal R}}
\newcommand{\uu}{\bm{p}}
\newcommand{\vv}{\bm{q}}
\newcommand{\Pmatrix}[1]{\begin{pmatrix} #1 \end{pmatrix}}
\newcommand{\wt}[1]{\widetilde{#1}}
\newcommand{\bm}[1]{\boldsymbol{#1}}
\newcommand{\diag}{\mathop{\rm diag}}
\newcommand{\bra}[1]{\langle #1\vert}
\newcommand{\ket}[1]{\vert #1\rangle}
\newcommand{\braket}[2]{\langle #1\vert #2\rangle}
\newcommand{\D}{D}
\newcommand{\W}{W}
\newcommand{\tr}{\mathop{\rm tr}}

\begin{titlepage}
\title{
\hfill\parbox{4cm}
{\normalsize KUNS-1753\\{\tt hep-th/0201177}}\\
\vspace{1cm}
{\bf Exact Results on Twist Anomaly}
}
\author{
Hiroyuki {\sc Hata}\thanks{
{\tt hata@gauge.scphys.kyoto-u.ac.jp}},
\
Sanefumi {\sc Moriyama}\thanks{
{\tt moriyama@gauge.scphys.kyoto-u.ac.jp}}
\
and
\
Shunsuke {\sc Teraguchi}\thanks{
{\tt teraguch@gauge.scphys.kyoto-u.ac.jp}}
\\[15pt]
{\it Department of Physics, Kyoto University, Kyoto 606-8502, Japan}
}
\date{\normalsize January, 2002}
\maketitle
\thispagestyle{empty}

\begin{abstract}
\normalsize
In vacuum string field theory, the sliver state solution has been
proposed as a candidate of a D-brane configuration.
Physical observables associated with this solution, such as its energy
density and the tachyon mass, are written in terms of the Neumann
coefficients. These observables, though vanish naively due to twist
symmetry, acquire non-vanishing values arising from their
singular behavior. Therefore, this phenomenon is called twist anomaly.
In this paper we present an analytical derivation of these physical
observables with the help of the star algebra spectroscopy.
We also identify in our derivation the origin of the twist anomaly in
the finite-size matrix regularization.
\end{abstract}
\end{titlepage}

\section{Introduction}
Vacuum string field theory (VSFT) \cite{RSZ1,RSZ2,RSZ3,VSFT} is a
candidate for the string field theory expanded around the tachyon
vacuum.
Since there are no open string excitations in the tachyon vacuum,
the VSFT conjecture claims that the string field theory around the
tachyon vacuum is obtained by replacing the BRST charge in cubic
string field theory (CSFT) \cite{Wit} by a purely ghost operator.
Although this sounds persuasive, we would like to know whether this
conjecture is true.
Hence, it is very important to construct the D25-brane solution
explicitly in VSFT and investigate the energy density of the
solution and the fluctuation spectrum around it.

A translationally and Lorentz invariant classical solution of VSFT
has been found in \cite{HatKaw} following earlier investigations
\cite{KosPot,RSZ2}.
Because this solution is believed to correspond to a D25-brane, it is
expected that the energy density of the solution is equal to the
D25-brane tension and the fluctuation spectrum around the solution
reproduces that of the ordinary open string theory.
To investigate the energy density and the fluctuation spectrum of the
solution, in \cite{HatKaw} a tachyon wave function was constructed
and the mass spectrum and the potential height are given in terms of
the Neumann coefficients expressing the three-string interactions.
By the numerical study of \cite{HatKaw,HatMor}, it was found that,
although the tachyon mass is correctly reproduced, contrary
to the expectation, the energy density is about twice the D25-brane
tension.

Furthermore, in \cite{HatMor} they evaluated these physical
observables by using various identities among the Neumann
coefficients. The result is as follows.
A Neumann coefficient matrix has a twist symmetry,
which causes the degeneracy of its eigenvalues.
Due to this degeneracy, the main constituents of the physical
observables seem to vanish naively.
However, actually in the numerical analysis, the observables acquire
non-vanishing values, since the expressions of the physical observables
behave singularly at the end of the eigenvalue distribution of the
Neumann coefficient \cite{HatMor,MooTay,BCFT}.
This is interpreted as a breakdown of the twist symmetry and hence
this phenomenon is called ``twist anomaly''.
Since the calculations in \cite{HatKaw,HatMor} have been done only by
numerical method, it should be very interesting if we could evaluate the
twist anomaly exactly.
For such analytic calculation, we have to solve the eigenvalue problem
of the Neumann coefficient matrices.

On the other hand, these observables in VSFT are computed exactly by
using boundary conformal field theory (BCFT) technique
\cite{BCFT}.
Using BCFT, they showed that the tachyon mass squared is exactly equal
to $-1/\ap$ and the ratio of the energy density to the D25-brane
tension is given by $(\pi^2/3)[16/(27\ln 2)]^3\simeq
2.05\cdots$, instead of the integer $2$.
Since the correspondence between these geometric method of BCFT and
the algebraic one using the Neumann coefficients is still not explicit,
it is worth repeating exact calculations by the algebraic method.

Recently, an elegant paper \cite{spec} has appeared, which solves the
eigenvalue problem of the Neumann coefficient matrix (star algebra
spectroscopy). In this paper,
using the results of \cite{spec}, we compute analytically the
observables such as the tachyon mass and the energy density by the
algebraic method to obtain the same results as those by the BCFT
method. In our analysis we adopt the regularization by cutting off the
size of the Neumann coefficient matrices to finite size ones.
In particular, we clarify the origin of the twist anomaly in this
regularization.
Other applications of the star algebra spectroscopy are found in
\cite{Oku2,Oku3,Okuda}.

The organization of the rest of this paper is follows.
In the next section, we review the classical solution of the VSFT
action, physical observables of this solution and their interpretation
as twist anomaly.
In sec.\ 3, we introduce the finite-size matrix regularization of the
Neumann coefficient matrices and calculate the tachyon mass by using
this regularization.
Sec.\ 4 is devoted to the evaluation of the energy density of the
solution.
In the final section, we summarize the paper and discuss some future
problems.
In appendix A and B, we present some technical details used in the
text.

\section{VSFT and its observables}
In this section, we shall review how the physical observables are
written in terms of the Neumann coefficients.
For this purpose, we shall first summarize the VSFT action and the
properties of the Neumann coefficients, and then proceed to the
construction of a classical solution and observables associated
with the solution.

\subsection{VSFT action and the Neumann coefficients}
The action of VSFT is given by \cite{RSZ1,RSZ2,VSFT}
\begin{equation}
\calS[\Phi]=-K\left(\frac{1}{2}\Phi\cdot\calQ\Phi
+\frac{1}{3}\Phi\cdot(\Phi*\Phi)\right),
\label{eq:SV}
\end{equation}
where $K$ is a constant and $\calQ$ is the purely ghost BRST operator
around the tachyon vacuum.
The star product $*$ is the same as in ordinary CSFT and defined
through the three-string vertex,
\begin{align}
\ket{V}_{123}&=\exp\left(-\frac{1}{2}\bm{A}^\dagger C\calM_3\bm{A}
-\bm{A}^\dagger\bm{V}
-\frac{1}{2}V_{00}(A_0)^2+(\mbox{ghost part})\right)\vac_{123}\nn\\
&\hspace{35ex}
\times(2\pi)^{26}\delta^{26}(p_1+p_2+p_3),
\label{eq:V}
\end{align}
with various quantities defined by $(n,m\ge 1)$
\begin{align}
&\bm{A}=\Pmatrix{a^{(1)}_n\\a^{(2)}_n\\a^{(3)}_n},
\quad
A_0=\Pmatrix{a^{(1)}_0\\a^{(2)}_0\\a^{(3)}_0},
\quad
\calM_3=\Pmatrix{M_0&M_+&M_-\\ M_-&M_0&M_+\\ M_+&M_-&M_0},
\nn\\
&\bm{V}=\Pmatrix{\bm{v}_0&\bm{v}_+&\bm{v}_-\\
                 \bm{v}_-&\bm{v}_0&\bm{v}_+\\
                 \bm{v}_+&\bm{v}_-&\bm{v}_0}\!A_0,
\quad V_{00}=\frac12\ln\left(\frac{3^3}{2^4}\right),
\quad C_{nm}=(-1)^n\delta_{nm}.
\label{eq:newquantities}
\end{align}
The matter oscillator $a^{(r)\mu}_n$ ($n\ge 1$) is normalized to
satisfy the commutation relation,
\begin{equation}
[a^{(r)\mu}_n,a^{(s)\nu\dagger}_m]
=\eta^{\mu\nu}\delta_{nm}\delta^{rs},
\label{CR}
\end{equation}
and $a^{(r)}_0$ is related to the center-of-mass momentum of the
string $r$, $p_r=-i\p/\p x_r$, by $a^{(r)}_0=\sqrt{2}\,p_r$
(we are adopting the convention of $\ap=1$).
We follow the notations of \cite{HatKaw,HatMor}.

The Neumann coefficients $M_0$, $M_\pm$, $\bm{v}_0$ and $\bm{v}_\pm$
satisfy the following linear relations:
\begin{align}
&M_0+M_++M_-=1,\\
&\bm{v}_0+\bm{v}_++\bm{v}_-=0
\label{eq:v+v+v=0}\\
&CM_0C=M_0,\quad C\bm{v}_0=\bm{v}_0,\\
&CM_\pm C=M_\mp,\quad C\bm{v}_\pm=\bm{v}_\mp.
\label{Linear}
\end{align}
It is convenient to introduce a new twist-odd matrix $M_1$
and vector ${\bm v}_1$,
\begin{align}
&\Mo=M_+-M_-,\\
&\vo=\bm{v}_+-\bm{v}_-,\\
&C\Mo C=-\Mo,\quad C\vo=-\vo,\label{CM1C}
\end{align}
and regard twist-even Neumann coefficients $\Me$ and $\ve$ and
twist-odd ones $\Mo$ and $\vo$ as independent.
Then, these Neumann coefficients are also known to satisfy the
following non-linear identities \cite{GroJev1,GroJev2,Kis}:
\begin{align}
&[\Me,\Mo]=0,\label{[M,M]}\\
&\Mo^2=(1-\Me)(1+3\Me),\label{MoOfMe}\\
&3(1-\Me)\ve+\Mo\vo=0,\label{Mv}\\
&3\Mo\ve +(1+ 3\Me)\vo=0,\label{Mv2}\\
&\frac94\,\ve^2+\frac34\vo^2=2\,V_{00}.\label{vv}
\end{align}

\subsection{Classical solution, tachyon  mode and observables}
The equation of motion of VSFT is given by
\begin{equation}
\calQ\Psic+\Psic*\Psic=0.
\label{eq:eqmot}
\end{equation}
Adopting the Siegel gauge for $\Psi_c$, $\ket{\Psi_c}=b_0\ket{\phic}$,
and the squeezed state ansatz for $\ket{\phic}$
\cite{KosPot,RSZ2,HatKaw}:
\begin{equation}
\ket{\phic}=\calNc\exp\biggl(
-\frac12\sum_{n,m\ge 1}a^\dagger_n(CT)_{nm}a^\dagger_m
+\sum_{n,m\ge 1}c^\dagger_n(C\wt{T})_{nm}b^\dagger_m\biggr)\vac ,
\label{eq:phic}
\end{equation}
the equation of motion (\ref{eq:eqmot}) is satisfied by choosing the
real matrix $T_{nm}$ as
\begin{equation}
T=\frac{1}{2 M_0}\left(1+M_0-\sqrt{(1-M_0)(1+3M_0)}\right) ,
\label{eq:T-}
\end{equation}
and the normalization factor $\calNc$ as
\begin{align}
\calNc=-\left[\det(1-T\calM)\right]^{13}
[\det(1-\wt{T}\wt{\calM})]^{-1},
\label{eq:calNc}
\end{align}
with
\begin{equation}
\calM=\Pmatrix{M_0&M_+\\ M_-&M_0}.
\end{equation}
The Neumann coefficients with a tilde are those for the ghost
part.
The form of the BRST operator $\calQ$ is uniquely fixed by
the requirement of the existence of the solution under the above
assumption \cite{HatKaw,GRSZ,Oku1,Oku2}.
The identification of the state (\ref{eq:phic}) with $T$ given by
(\ref{eq:T-}) as the sliver state \cite{sliver,RSZ2} has recently been
proved by \cite{Okuda}.

To see whether this solution corresponds to a D$25$-brane, we have to
examine whether the fluctuation spectrum and the energy density of
this solution give the expected ones.
As the first step of the examination of the mass spectrum, we solve
the linearized equation of motion for the tachyon wave function
$\Phi_t$,
\begin{equation}
\calQB\Phi_t\equiv\calQ\Phit+\Psic*\Phit+\Phit*\Psic=0,
\label{eq:QBPhit}
\end{equation}
where $\calQB$ is the BRST operator for fluctuations around $\Psic$.
A plausible choice for the tachyon fluctuation mode
$\ket{\Phi_t}=b_0\ket{\phit}$ in the Siegel gauge was proposed in
\cite{HatKaw}:
\begin{equation}
\ket{\phit}=\frac{\calNt}{\calNc}
\exp\biggl(-\sum_{n\ge 1}t_n a^\dagger_n a_0\biggr)\ket{\phic}.
\label{eq:phit}
\end{equation}
It was found that the linearized equation of motion (\ref{eq:QBPhit})
is satisfied if the vector $\bm{t}$ is given by
\begin{equation}
\bm{t}=3(1+T)(1+3M_0)^{-1}\ve ,
\label{eq:t}
\end{equation}
and the center-of-mass momentum $p^\mu=a_0^\mu/\sqrt{2}$ satisfies the
following on-shell condition:
\begin{equation}
p^2=-m_t^2\equiv\frac{\ln 2}{G}.
\label{onshell}
\end{equation}
Here $G$ is given in terms of the Neumann coefficients as follows:
\begin{align}
&G=2V_{00}+(\bm{v}_--\bm{v}_+,\bm{v}_+-\bm{v}_0)(1-T\calM)^{-1}T
\Pmatrix{\bm{v}_+-\bm{v}_-\\ \bm{v}_--\bm{v}_0}\nn\\
&\hspace{2cm}
+2(\bm{v}_--\bm{v}_+,\bm{v}_+-\bm{v}_0)
(1-T\calM)^{-1}\Pmatrix{0\\\bm{t}}
+(0,\bm{t})\calM(1-T\calM)^{-1}\Pmatrix{0\\ \bm{t}}.
\label{G}
\end{align}
The normalization factor $\calNt$ for $\Phit$ in (\ref{eq:phit})
is determined to be
\begin{equation}
\calNt=\frac{1}{\sqrt{KG}}
\left[\det(1-T^2)\right]^{13/2}\bigl[\det(1-\wt{T}^2)\bigr]^{-1/2}
\exp\Bigl(\bm{t}\,(1+T)^{-1}\bm{t}\,\mt^2\Bigr),
\label{eq:calNt}
\end{equation}
from the requirement that $\Phit$ has a canonical kinetic term:
\begin{equation}
\frac{K}{2}\Phit\cdot\calQB\Phit\underset{p^2\sim-\mt^2}{\sim}
-\frac12(p^2+\mt^2) .
\end{equation}

As the first test of the present classical solution $\Psic$ and tachyon
mode $\Phit$, we have to check whether the tachyon mass $m_t^2$
(\ref{onshell}) reproduces the correct value of $-1$.
It was found numerically in \cite{HatKaw} that, by truncating the
infinite-size Neumann coefficient matrices into finite but large
ones, the quantity $G$ actually gives the expected value $\ln 2$.
Later in \cite{HatMor}, an interesting interpretation of $G$ (and other
observables in VSFT) was presented.
They found that this quantity $G$
(\ref{G}) vanishes identically, if we naively use the various
non-linear relations among the Neumann coefficients,
(\ref{[M,M]})--(\ref{Mv}).
The vanishing of $G$ can be ascribed to that the eigenvalues of
the matrix $\Me$ are doubly degenerate between twist-even and odd
eigenvectors.
This phenomenon that the quantity $G$ which vanishes naively due to
twist symmetry can actually have non-vanishing value is called
``twist anomaly'' in \cite{HatMor}.
Further in \cite{HatMor}, from the numerical analysis they found that
this paradox emerges because the expression of $G$ (\ref{G}) is
singular at $\Me=-1/3$ (actually the eigenvalue distribution of
$\Me$ ranges between $-1/3$ and $0$ \cite{HatMor,MooTay,spec}).
Due to this singularity, the expression (\ref{G}) takes a form of the
difference of two divergent quantities, and hence a careful treatment
is necessary. On the basis of numerical analysis, the following
calculation rules leading to correct values has been proposed:
\renewcommand{\arraystretch}{1.3}
\begin{table}[htbp]
\begin{center}
\begin{tabular}[b]{|c|c|c|c|c|}
\hline
$1/\sqrt{1+3M_0}$&$\Mo$&$\ve$&$\vo$&$\bm{t}$\\
\hline\hline
$1$&$-1$&$0$&$1$&$1$\\
\hline
\end{tabular}
\caption{Degrees of singularity for various quantities.}
\label{tab:degree}
\end{center}
\end{table}
\renewcommand{\arraystretch}{1}
\begin{itemize}
\item
We assign each Neumann coefficient matrix its degree of singularity
as given in table \ref{tab:degree}. This assignment is compatible with
the nonlinear relations (\ref{MoOfMe})--(\ref{Mv2}).
\item
We laurent-expand the quantity such as $G$ around the singular point
$M_0=-1/3$.
\item
For the terms with degrees of singularity less than three, we can
freely use the nonlinear relations (\ref{[M,M]})--(\ref{vv}).
\item
For the most singular terms with degree three, we treat them as they
stand. However, we are allowed to use the nonlinear relation
(\ref{Mv}) to express $\ve$ in terms of $\vo$;
$\ve=-(1/3)(1-\Me)^{-1}\Mo\vo$.
\end{itemize}
Using this rule, we find that the quantity $G$ (\ref{G}) is
simplified into the following form:
\begin{equation}
G=-\frac{9\sqrt{3}}{32}\,\vo\biggl(
\Mo\frac{1}{(1+3\Me)^{3/2}}\Mo-\frac{1-\Me}{\sqrt{1+3\Me}}
\biggr)\vo.
\label{eq:Gdeform}
\end{equation}
In the next section we shall regularize this indefinite expression
properly and evaluate it analytically.

Now let us turn to the energy density of the classical solution.
First, the energy density $\calEc$ of the solution $\Psic$ is given by
\begin{equation}
\calEc=-\frac{\calS[\Psic]}{V_{26}}
=\frac{K}{6}
\Biggl(\frac{\left[\det(1-T\calM)\right]^2}{\det(1-T^2)}\Biggr)^{13}
\Biggl(\frac{
[\det(1-\wt{T}\wt{\calM})]^2}{\det(1-\wt{T}^2)}\Biggr)^{-1},
\label{eq:calEc}
\end{equation}
For comparing $\calEc$ with the D$25$-brane tension $T_{25}$, let us
calculate the latter. It is given in the
present convention of $\ap=1$ by $T_{25}=1/(2\pi^2\go^2)$ with $\go$
being the open string coupling constant defined as the three-tachyon
on-shell amplitude.
Using the tachyon wave function $\Phit$ (\ref{eq:phit}),
$\go$ is given by
\begin{align}
\go&=K\,\Phit\cdot(\Phit *\Phit)
\Bigr\vert_{p_1^2=p_2^2=p_3^2=-\mt^2}\nn\\
&=K\calNt^3\,\left[\det(1-T\calM_3)\right]^{-13}
\det(1-\wt{T}\wt{\calM}_3)
\,\exp\biggl\{
-\frac12\bm{V}(1-T\calM_3)^{-1}TC\bm{V}
\nn\\
&\qquad\qquad\qquad
+\bm{V}(1-T\calM_3)^{-1}\bm{t}A_0
-\frac12 A_0\bm{t}\calM_3(1-T\calM_3)^{-1}\bm{t}A_0
-\frac12 V_{00}(A_0)^2
\biggr\}.
\label{eq:KphitphitphitV}
\end{align}

{}From (\ref{eq:calEc}), (\ref{eq:KphitphitphitV}) and
(\ref{eq:calNt}),
we find the expression for the ratio $\calEc/T_{25}$:
\begin{equation}
\frac{\calEc}{T_{25}}=\frac{\pi^2}{3G^3}\exp(6\mt^2H),
\label{eq:ratio}
\end{equation}
where $H$ is given by
\begin{align}
H=&-\frac{2}{(A_0)^2}
\left[-\frac12\bm{V}(1-T\calM_3)^{-1}TC\bm{V}
+\bm{V}(1-T\calM_3)^{-1}\bm{t}A_0
-\frac12 A_0\bm{t}\calM_3(1-T\calM_3)^{-1}\bm{t}A_0
\right]\nn\\[7pt]
&\quad+\bm{t}(1+T)^{-1}\bm{t}+V_{00}.
\label{eq:totalH}
\end{align}
Following the above calculation rules, we can simplify
(\ref{eq:totalH}) into
\begin{align}
H=&\frac{\sqrt{3}}{4}\vo\frac{1}{\sqrt{1+3\Me}}\biginv\vo
-\frac{3\sqrt{3}}{8}\vo\frac{1}{\sqrt{1+3\Me}}\Mo
\frac{1}{\sqrt{1+3\Me}}\biginv\frac{1}{\sqrt{1+3\Me}}\Mo\vo\nn\\
&\quad-\frac{9\sqrt{3}}{16}\vo\Mo\frac{1}{1+3\Me}\biginv
\frac{1}{\sqrt{1+3\Me}}\Mo\vo
+\frac{9\sqrt{3}}{16}\vo\Mo\frac{1}{(1+3\Me)^{3/2}}\Mo\vo\nn\\
&\quad-\frac{\sqrt{3}}{16}\vo\frac{(1+3\Me)^{3/2}}{5-\Me}\vo ,
\label{eq:simpleH}
\end{align}
where $\biginv$ is defined by
\begin{equation}
\biginv =\biggl(
1+\frac14\Mo\frac{1}{\sqrt{1+3\Me}}\Mo\frac{1}{\sqrt{1+3\Me}}
\biggr)^{-1}.
\end{equation}
Analytic evaluation of $H$ will be given in sec.\ 4.

\section{Tachyon mass}
Let us proceed to evaluating the tachyon mass or the quantity
$G$ analytically.
{}From the numerical analysis in \cite{HatKaw,HatMor} we know that the
quantity $G$ gives the expected value $\ln 2$ to high precision.
We shall show analytically that this is exact.
The analysis in this section will also reveal how the breakdown of twist
symmetry brought about by regularization makes $G$ non-vanishing.

\subsection{Finite-size matrix regularization}
Before proceeding to evaluating the tachyon mass exactly, we
shall explain how to regularize the expression (\ref{eq:Gdeform}) of
$G$ which is indefinite due to the eigenvalue $-1/3$ of $\Me$.
For this purpose we shall first present expressions of the Neumann
coefficient matrices in terms of simpler quantities.

In \cite{spec} it was found that the Neumann coefficient matrices
$M_0$ and $M_1$ are related to another simpler matrix $\K$, which is
the matrix representation of the Virasoro algebra, $\K=L_1+L_{-1}$, by
\begin{align}
M_0&=-\frac{1}{1+2\cosh(\K\pi/2)},\label{M0ofK1}\\
M_1&=\frac{2\sinh(\K\pi/2)}{1+2\cosh(\K\pi/2)}.\label{M1ofK1}
\end{align}
The matrix representation of $\K$ can be read off from the
operation of the Virasoro algebra:
\begin{equation}
\bigl(\K\bigr)_{nm}
=-\sqrt{(n-1)n}\,\delta_{n-1,m}-\sqrt{n(n+1)}\,\delta_{n+1,m} .
\label{K1nm}
\end{equation}
Note that $\K$ is symmetric and twist-odd:
\begin{equation}
\K^T =\K,\quad  C\K C=-\K .
\label{eq:KT=K}
\end{equation}

For the vectors $\ve$ and $\vo$, we have the following convenient
expressions:
\begin{align}
\ve&=-\frac13(1+3M_0)\q,\label{v0}\\
\vo&=M_1\q,\label{v1}
\end{align}
with a new vector $\q$ defined as
\begin{equation}
u_n=\frac{1}{\sqrt{n}}\cos\left(\frac{n\pi}{2}\right)=
\begin{cases}
\ds\frac{(-1)^{n/2}}{\sqrt{n}}&\text{$n$: even}\\0&\text{$n$: odd}
\end{cases}
.\label{eq:u}
\end{equation}
Proof of (\ref{v0}) and (\ref{v1}) is given in appendix
\ref{app:vector}.

Now we can express $G$ (\ref{eq:Gdeform}) in terms of the simple
matrix $\K$ and vector $\q$.
Adopting the calculation rules in \cite{HatMor}, which we summarized
in sec.\ 2.2, all we have to do is to concentrate on the most singular
term.
Namely, we assign $\K$ and $\q$ their degrees of singularity $-1$ and
$2$, respectively, and laurent-expand $\Me$, $\Mo$ and $\vo$ in $G$
with respect to $\K$ around the singularity $\K=0$ corresponding to
$\Me=-1/3$:
\begin{align}
1+3M_0&\simeq\frac{\pi^2}{12}\K^2,\label{M0simK1}\\
M_1&\simeq\frac{\pi}{3}\K ,\label{M1simK1}\\
\vo&\simeq\frac{\pi}{3}\K\q. \label{eq:vo=Ku}
\end{align}
Keeping only the terms with degree three, the quantity $G$ can be
expressed as\footnote{
Since the RHS of (\ref{GqK}) manifestly vanishes by naive treatment,
we do not need to add it the correction term $G_{\rm reg}$ with degree
less than three (see sec.\ 3.2 of \cite{HatMor}).
}
\begin{equation}
G=\frac{\pi}{4}\biggl(
\q^T \K\Kabsinv\K\q-\q^T \K\K\Bigl(\Kabsinv\Bigr)^3\K\K\q
\biggr).
\label{GqK}
\end{equation}
The present expression of $G$ is indefinite as ever, and we have to
regularize it.
In the numerical analysis of the original expression of $G$
(\ref{eq:Gdeform}), we adopted the level truncation, namely we cut-off
all the infinite dimensional matrices into $L\times L$ ones.
While $\Me$ without regularization has the eigenvalue $-1/3$
\cite{HatMor,MooTay,spec}, in the level truncation the lowest
eigenvalue of $\Me$ is lifted from $-1/3$.
Hence, this level truncation serves as a good regularization.

In this finite-size regularization, we have truncated the
infinite-size matrices $\Me$ and $\Mo$ into those of the same size
$L$.
Therefore, the regularized version of the matrix $\K^2$ in the
denominator of (\ref{GqK}) should be the truncation of the square of
the infinite dimensional matrix, $(\K)^2\bigr|_L$, since this $\K^2$
originates from $1+3\Me$ (\ref{M0simK1}).
On the other hand, the regularized version of $\K$ in the numerator of
(\ref{GqK}), which comes from $\Mo$ (\ref{M1simK1}), should simply be
the truncation of $\K$ itself, $\K\bigr|_L$.

\subsection{Matrix representation}
In this subsection, let us study how the degeneracy of the eigenvalues
between twist-even and odd eigenvectors is lifted due to finite-size
matrix regularization.
Here we shall change rows and columns of a generic matrix $M$.
We shall bring rows with odd indices into the upper side and ones with
even indices into the lower side, and repeat a similar manipulation
for columns:
\begin{equation}
M=\Pmatrix{M_{oo}&M_{oe}\\M_{eo}&M_{ee}}.
\label{Moe}
\end{equation}
Since the matrix $\K$ is twist-odd, $C\K C=-\K$, its
diagonal blocks vanish in this representation.
\begin{equation}
\K\Bigr|_L=\Pmatrix{0&(\K)_{oe}\\(\K)_{eo}&0}.
\label{Ktruncated}
\end{equation}
Although the off-diagonal blocks are not symmetric matrices by
themselves, we can give them a useful decomposition.
First let us consider the case that the truncation level $L$ is an
even number $2\ell$.
After squaring (\ref{Ktruncated}), we find the diagonal block
symmetric.
Therefore, we can diagonalize it:
\begin{equation}
\Bigl(K_1\Bigr|_{2\ell}\Bigr)^2
=\Pmatrix{\Pm_{2\ell}\Lambda_{2\ell}^2\Pm_{2\ell}^T&0\\
0&\Qm_{2\ell}\Lambda_{2\ell}^2\Qm_{2\ell}^T} ,
\label{K|2l^2}
\end{equation}
where the diagonal matrix $\Lambda_{2\ell}$ and the orthogonal
matrices $\Pm_{2\ell}$ and $\Qm_{2\ell}$ are all $\ell\times\ell$
ones.
Note here that the eigenvalues $\kappa$ of the odd-odd sector and
even-even one degenerate because their eigenvalue equations are
identical:
\begin{align}
\det\Bigl((\K)_{oe}(\K)_{eo}-\kappa^2 1\Bigr)=0
\quad\Leftrightarrow\quad
\det\Bigl((\K)_{eo}(\K)_{oe}-\kappa^2 1\Bigr)=0.
\end{align}
The eigenvalue distribution of finite-size regularized matrix $\K$ was
analyzed in \cite{spec}. They found that the spacing between the
nearest eigenvalues is independent of $\kappa$ and given by
\begin{align}
\Delta|\kappa|=\frac{2\pi}{\ln L} .
\label{spacing}
\end{align}
Assuming that the spacing (\ref{spacing}) applies also to two adjacent
eigenvalues with opposite sign\footnote{
Numerical analysis of the smallest solution $\kappa$ to (6.3) in
\cite{spec} supports this assumption to high precision.
}
and using the facts that the eigenvalues $\kappa$ and $-\kappa$ are
always paired and that there exits no zero eigenvalue in the case
$L=2\ell$, the eigenvalues of $\K\bigr|_{2\ell}$, which are the
diagonal elements of $\Lambda_{2\ell}$, are given by
\begin{equation}
\bigl(\Lambda_{2\ell}\bigr)_{nn}
=\frac{2\pi}{\ln L}\biggl(n-\frac12\biggr)\equiv\kappa_{n-\frac12} .
\label{Lambda2l}
\end{equation}
Note that the degeneracy in (\ref{K|2l^2}) always occurs for any
finite-size matrix $M$ which is symmetric $M^T=M$ and twist-odd
$CMC=-M$.

Using the expression (\ref{K|2l^2}) of
$\bigl(\K\bigr|_{2\ell}\bigr)^2$,
we find a useful decomposition for $\K$.
\begin{equation}
K_1\Bigr|_{2\ell}
=\Pmatrix{0&\Pm_{2\ell}\Lambda_{2\ell}\Qm_{2\ell}^T\\
\Qm_{2\ell}\Lambda_{2\ell}\Pm_{2\ell}^T&0}.
\label{K|2l}
\end{equation}
This decomposition is unique up to the overall sign.
This can be seen by counting the degrees of freedom in the matrix.
Each of the $\ell\times\ell$ orthogonal matrices $\Pm_{2\ell}$ and
$\Qm_{2\ell}$ has $\ell(\ell-1)/2$ degrees of freedom, and the
diagonal matrix $\Lambda_{2\ell}$ has $\ell$ degrees of freedom.
Therefore, the number of degrees of freedom in the RHS of (\ref{K|2l})
is equal to $\ell(\ell-1)/2\times 2+\ell=\ell^2$, which agrees with
that of $(\K)_{oe}=\left[(\K)_{eo}\right]^T$ on the RHS of
(\ref{Ktruncated}).
Since the expression (\ref{K|2l}) has the same number of degrees of
freedom as in the original matrix (\ref{Ktruncated}), we see that the
decomposition is unique.

The same argument holds for the case that the truncation level $L$ is
an odd number $2\ell+1$, except that in this case we encounter
rectangular matrices and a careful analysis is necessary.
In this case the square of the matrix $\K$ is
\begin{equation}
\Bigl(K_1\Bigl|_{2\ell+1}\Bigr)^2
=\Pmatrix{\Pm_{2\ell+1}\diag(\Lambda_{2\ell+1}^2,0)\Pm_{2\ell+1}^T&0\\
0&\Qm_{2\ell+1}\Lambda_{2\ell+1}^2\Qm_{2\ell+1}^T}.
\label{K|2l+1^2}
\end{equation}
Here $\Lambda_{2\ell+1}$ and $\Qm_{2\ell+1}$ are $\ell\times\ell$
matrices, while $\Pm_{2\ell+1}$ is a $(\ell+1)\times(\ell+1)$ matrix.
Since both the odd-odd block (an $(\ell+1)\times(\ell+1)$ matrix)
and the even-even one (an $\ell\times\ell$ matrix) have the same
rank, the former one should have an extra zero eigenvalue.
The diagonal matrix $\Lambda_{2\ell+1}$ is now given as
\begin{equation}
\bigl(\Lambda_{2\ell+1}\bigr)_{nn}=\frac{2\pi}{\ln L}n\equiv\kappa_n.
\label{Lambda2l+1}
\end{equation}
The eigenvalue spectrum (\ref{Lambda2l+1}) can be understood from the
spacing (\ref{spacing}) and the fact that we have a zero eigenvalue in
the present case.
Hence, the matrix $\K\bigr|_{2\ell+1}$ by itself reads
\begin{equation}
K_1\Bigr|_{2\ell+1}
=\Pmatrix{0&\Pm_{2\ell+1}(\Lambda_{2\ell+1},\bm{0})^T\Qm_{2\ell+1}^T\\
\Qm_{2\ell+1}(\Lambda_{2\ell+1},\bm{0})\Pm_{2\ell+1}^T&0},
\label{K|2l+1}
\end{equation}
where $(\Lambda_{2\ell+1},\bm{0})$ is an $\ell\times(\ell+1)$ matrix
with vanishing $(\ell+1)$-th column.

Now let us return to the expression (\ref{GqK}) of $G$.
As we explained in the previous subsection, the regularized version of
$\K^2$ in the denominator stands for $\bigl(\K\bigr)^2\bigr|_L$,
namely the truncation of the square of the original infinite
dimensional matrix $\K$.
We see here how the breakdown of the twist symmetry happens in
this regularization.
In the original representation before changing rows and
columns into (\ref{Moe}), the difference between truncating before
squaring, $\bigl(\K\bigr|_L\bigr)^2$, and truncating after squaring,
$\bigl(\K\bigr)^2\bigr|_L$, appears only at the last $(L,L)$
component.
As seen from (\ref{K1nm}), the last component of
$\bigl(\K\bigr|_L\bigr)^2$ is $(L-1)L$, while that of
$\bigl(\K\bigr)^2\bigr|_L$ is $(L-1)L+L(L+1)=2L^2$.
Therefore, when $L=2\ell$, since the last component in the original
representation belongs to the even-even block, the odd-odd block of
the matrix $\bigl(\K\bigr)^2\bigr|_{2\ell}$ should be the same as
that of $\bigl(\K\bigr|_{2\ell}\bigr)^2$.
As for the even-even block of $\bigl(\K\bigr)^2\bigr|_{2\ell}$, it can
be read off from another matrix $\bigl(\K\bigr|_{2\ell+1}\bigr)^2$.
Note that, while  the odd-odd block of this matrix
$\bigl(\K\bigr|_{2\ell+1}\bigr)^2$ is enlarged to
$(\ell+1)\times(\ell+1)$, its even-even block is of the same
size $\ell\times\ell$ as the even-even block of
$\bigl(\K\bigr)^2\bigr|_{2\ell}$.
Since the last $(2\ell+1,2\ell+1)$ component of
$\bigl(\K\bigr|_{2\ell+1}\bigr)^2$ belongs to the odd-odd block,
the even-even block of this matrix is the same as that
of $\bigl(\K\bigr)^2\bigr|_{2\ell}$, which is of our interest.
Therefore the final expression of $\bigl(\K\bigr)^2\bigr|_{2\ell}$ is
given by
\begin{align}
(K_1)^2\Bigr|_{2\ell}
=\Pmatrix{\Bigl[\bigl(\K\bigr|_{2\ell}\bigr)^2\Bigr]_{oo}&0\\
0&\Bigl[\bigl(\K\bigr|_{2\ell+1}\bigr)^2\Bigr]_{ee}}
=\Pmatrix{\Pm_{2\ell}\Lambda_{2\ell}^2\Pm_{2\ell}^T&0\\
0&\Qm_{2\ell+1}\Lambda_{2\ell+1}^2\Qm_{2\ell+1}^T}.
\label{K^2|2l}
\end{align}
Note that the degeneracy of eigenvalues between the odd-odd block and
the even-even one is in fact lifted in the regularized expression
(\ref{K^2|2l}).
In the case $L=2\ell+1$, a similar argument gives
\begin{align}
(K_1)^2\Bigr|_{2\ell+1}
=\Pmatrix{\Pm_{2\ell+2}\Lambda_{2\ell+2}^2\Pm_{2\ell+2}^T&0\\
0&\Qm_{2\ell+1}\Lambda_{2\ell+1}^2\Qm_{2\ell+1}^T}.
\label{K^2|2l+1}
\end{align}

\subsection{Evaluation of $G$}
Having seen in the previous two subsections how we should regularize
the singularity in the physical observables and how the
degeneracy of the eigenvalues due to twist symmetry is lifted in this
regularization,
let us proceed to evaluating $G$ analytically.
For simplicity, in the following we shall take the truncation level
$L$ to be an even number.
Noting that
\begin{align}
(\K\q)_n&=\begin{cases}
1&\mbox{$n$=1}\\0&\mbox{otherwise}
\end{cases},
\label{Ku}
\\
(\K\K\q)_n&=\begin{cases}
-\sqrt{2}&\mbox{$n$=2}\\0&\mbox{otherwise}
\end{cases},
\end{align}
which can be easily seen from (\ref{K1nm}) and (\ref{eq:u}), we can
further simplify the expression (\ref{GqK}) into
\begin{align}
G&=\frac{\pi}{4}\biggl(
\Kabsinv-\K\Bigl(\frac{1}{\Kabs}\Bigr)^3\K\biggr)[1,1]
\nn\\
&=\frac{\pi}{4}\biggl(
\frac{1}{\sqrt{\K^2}}[1,1]-2\Bigl(\frac{1}{\Kabs}\Bigr)^3[2,2]
\biggr).
\label{GofK}
\end{align}
Here $M[n,m]$ for a generic matrix stands for its $(n,m)$ component
$M_{n,m}$.
The component indices in (\ref{Ku})--(\ref{GofK}) are those
in the original representation before changing to the representation
of (\ref{Moe}).

It is by no means an easy task to find one of the components of the
matrices in (\ref{GofK}) analytically.
However, since the eigenvalue problem of the matrix $\K$ is solved in
\cite{spec}, we can evaluate $G$ by using it.
The eigenvector $\bm{f}^{(\kappa)}$ of $\K$
corresponding to the eigenvalue $\kappa$,
\begin{equation}
\K\bm{f}^{(\kappa)}=\kappa\bm{f}^{(\kappa)} ,
\label{eq:Kf=kf}
\end{equation}
is given by (\ref{fcomp}).
We define the twist-odd eigenvectors $\uu_n$ and twist-even
eigenvectors $\vv_n$ of $\K^2$ as
\begin{align}
&\bigl(\uu_n\bigr)_m=\frac12
\Bigl(\bm{f}^{(\kappa_n)}+\bm{f}^{(-\kappa_n)}\Bigr)_{2m-1}
=\biggl(1,\frac{\kappa_n^2-2}{2\sqrt{3}},\cdots\biggr),
\label{pn}\\
&\bigl(\vv_n\bigr)_m=\frac12
\Bigl(\bm{f}^{(\kappa_n)}-\bm{f}^{(-\kappa_n)}\Bigr)_{2m}
=\biggl(-\frac{\kappa_n}{\sqrt{2}},
-\frac{\kappa_n^3-8\kappa_n}{12},\cdots\biggr),
\label{qn}
\end{align}
for $\kappa_n=2\pi n/\ln L$ with integer $n$ and similar ones
$\uu_{n-\frac12}$ and $\vv_{n-\frac12}$ for
$\kappa_{n-\frac12}=2\pi\left(n-\frac12\right)/\ln L$.
The matrices $\Pm_{2\ell}$, $\Pm_{2\ell+1}$, $\Qm_{2\ell}$ and
$\Qm_{2\ell+1}$ are expressed using these eigenvectors.
Introducing new symbols for these matrices to avoid cumbersome
subscripts, we have
\begin{align}
\Po&\equiv \Pm_{2\ell}=\bigl(
\overline{\uu}_{\frac12},\overline{\uu}_{\frac32},\cdots\bigr) ,
\label{Po}\\
\Pe&\equiv \Pm_{2\ell+1}\bigr|_{\mbox{\scriptsize zero-mode removed}}
=\bigl(\overline{\uu}_1,\overline{\uu}_2,\cdots\bigr),
\label{Pe}\\
\Qo&\equiv \Qm_{2\ell}=\bigl(
\overline{\vv}_{\frac12},\overline{\vv}_{\frac32},\cdots\bigr),
\label{Qo}\\
\Qe&\equiv \Qm_{2\ell+1}=\bigl(
\overline{\vv}_1,\overline{\vv}_2,\cdots\bigr),
\label{Qe}
\end{align}
where the vectors with a bar, $\overline{\uu}$ and $\overline{\vv}$,
denote the normalized ones of $\uu$ and $\vv$.
The subscripts H and I imply half-an-odd integer and integer
eigenvalues (in unit of $2\pi/\ln L$), respectively.
We have defined $\Pe$ as $\Pm_{2\ell+1}$ with the eigenvector
corresponding to the zero eigenvalue removed.

Using the expression of $1/\Kabs$,
\begin{equation}
\Kabsinv =\Pmatrix{
\Po\Lo^{-1}\Po^T & 0\\
0 & \Qe\Le^{-1}\Qe^T
} ,
\label{1/K}
\end{equation}
where $\Lo$ and $\Le$ are the diagonal matrices of eigenvalues
\begin{align}
&\Lo\equiv\Lambda_{2\ell}=\diag\bigl(\kappa_{n-\frac12}\bigr) ,
\\
&\Le\equiv\Lambda_{2\ell+1}=\diag\bigl(\kappa_{n}\bigr) ,
\end{align}
eq.\ (\ref{GofK}) is rewritten into
\begin{align}
G&=\frac{\pi}{4}\biggl(
\sum_n\Bigl(\bigl(\overline{\uu}_{n-\frac12}\bigr)_1\Bigr)^2
\frac{1}{\kappa_{n-\frac12}}
-\sum_n\Bigl(\bigl(\overline{\vv}_n\bigr)_1\Bigr)^2
\frac{2}{\kappa_n^3}
\biggr)\nn\\
&=\frac{\pi}{4}\biggl(
\sum_n\frac{1}{|\uu_{n-\frac12}|^2}\frac{1}{\kappa_{n-\frac12}}
-\sum_n\frac{(-\kappa_n/\sqrt{2})^2}{|\vv_n|^2}
\frac{2}{\kappa_n^3}\biggr),
\label{cancel}
\end{align}
where $|\bm{a}|$ denotes the norm of a vector $\bm{a}$.

The norm of the eigenvectors is derived in appendix \ref{app:norm}.
Especially for the vector in the finite $L$ regularization, the
norm is given by (\ref{norm}) with the delta function (\ref{delta}).
Hence, from the definition of our eigenvectors $\uu_{n-\frac12}$ and
$\vv_n$, (\ref{pn}) and (\ref{qn}), we find that their norms are
\begin{align}
|\uu_{n-\frac12}|^2
&=\frac{\delta(0)\sinh\bigl(\kappa_{n-\frac12}\pi/2\bigr)}
{\kappa_{n-\frac12}} ,\\
|\vv_n|^2
&=\frac{\delta(0)\sinh\bigl(\kappa_n\pi/2\bigr)}{\kappa_n} ,
\end{align}
with $\delta(0)=\ln L/(2\pi)$ in the finite $L$
regularization.
Therefore our expression for $G$ is reduced to
\begin{equation}
G=\frac{\pi}{4\delta(0)}\biggl(\sum_{n=1}^{L/2}
\frac{1}{\sinh\bigl(\kappa_{n-\frac12}\pi/2\bigr)}
-\sum_{n=1}^{L/2}\frac{1}{\sinh\bigl(\kappa_n\pi/2\bigr)}\biggr) .
\label{G=1/sinh}
\end{equation}
In the limit $L\to\infty$, we can replace $\sinh x$ by $x$ in
(\ref{G=1/sinh}) and finally obtain the desired result:
\begin{equation}
G=\frac{1}{2}\sum_{n=1}^\infty
\Bigl(\frac{1}{n-1/2}-\frac{1}{n}\Bigr)
=\ln 2 .
\label{Gseries}
\end{equation}

The reason why we have obtained a non-vanishing value of $G$ is that
the degeneracy of eigenvalues between odd-odd and even-even sectors is
lifted in the finite $L$ regularization as seen from (\ref{K^2|2l})
and (\ref{1/K}). As remarked below (\ref{Lambda2l}), the degeneracy is
a general property of twist-odd symmetric matrices. Therefore, the
phenomenon that a quantity such as $G$ vanishing  naively owing to the
degeneracy acquires a non-zero value was called twist anomaly in
\cite{HatMor}.
Note also that the non-vanishing result of (\ref{Gseries}) comes only
from infinitesimally small eigenvalues of order $1/\ln L$ in the limit
$L\to\infty$. This should be regarded as a precise expression of (3.17)
of \cite{HatMor} which has contribution only from the zero eigenvalue
$\kappa=0$ ($M_0=-1/3$).

\subsection{Properties of $\bm{P}$ and $\bm{Q}$}

In this subsection we shall derive a number of properties of the
matrices $\Po$, $\Pe$, $\Qo$ and $\Qe$ defined by
(\ref{Po})--(\ref{Qe}).
These properties are useful for systematic evaluation of the
observables. In the last part of this subsection we shall rederive
(\ref{Gseries}) for $G$ by using the properties.

First, as can be seen from the inner product formula
(\ref{inner}) (with $\sinh(\lambda\pi/2)$ approximated by
$\lambda\pi/2$), the vectors $\uu_n$ (\ref{pn}), $\vv_n$ (\ref{qn})
and their half-an-odd counterparts satisfy the orthogonality,
\begin{equation}
\uu_n\cdot\uu_m=\vv_n\cdot\vv_m=
\uu_{n-\frac12}\cdot\uu_{m-\frac12}=\vv_{n-\frac12}\cdot\vv_{m-\frac12}
=\frac{\ln L}{4}\,\delta_{n,m} .
\label{uuvv1}
\end{equation}
Corresponding to this fact, the three matrices $\Po$, $\Qo$ and $\Qe$
are orthogonal ones:
\begin{equation}
\calO^T\calO=\calO\calO^T =1 ,
\quad (\calO=\Po, \Qo, \Qe) .
\label{eq:O^TO=1}
\end{equation}
However, since the zero-mode is removed from the matrix $\Pe$
(\ref{Pe}), though we have
\begin{equation}
\Pe^T\Pe=1
\label{eq:Pe^TPe=1} ,
\end{equation}
the completeness relation $\Pe\Pe^T=1$ does not hold.

To derive the formulas associated with $\Pe$, let us consider
the products $\Po^T\Pe$ and $\Qo^T\Qe$. Their components are
calculated by using (\ref{inner}) to be given by\footnote{
Strictly speaking, the inner product $\uu_{n-\frac12}\cdot\uu_m$
between the $\ell$ component vector $\uu_{n-\frac12}$ and the
$(\ell+1)$ one $\uu_m$ is defined by removing the $(\ell+1)$-th
component of $\uu_m$. Equivalently, the matrix product $\Po^T\Pe$
should be understood to imply $\Po^T(1_{\ell\times\ell},\bm{0})\Pe$.
}
\begin{align}
&\left(\Po^T\Pe\right)_{n,m}
=\frac{4}{\ln L}\,\uu_{n-\frac12}\cdot\uu_m
=-\left(n-\frac12\right)\D_{n,m} ,
\label{eq:PoPe}
\\
&\left(\Qo^T\Qe\right)_{n,m}
=\frac{4}{\ln L}\,\vv_{n-\frac12}\cdot\vv_m
=-\D_{n,m}\,m ,
\label{eq:QoQe}
\end{align}
where $D_{n,m}$ is defined by
\begin{equation}
\D_{n,m}=\frac{2}{\pi}\frac{(-1)^{n+m}}{(n-1/2)^2-m^2} .
\label{eq:D}
\end{equation}
Then, the following formula is an immediate consequence of
(\ref{eq:PoPe}) and (\ref{eq:QoQe}):
\begin{equation}
\Po^T\Pe\Le=\Lo\Qo^T\Qe .
\label{eq:formulaone}
\end{equation}
As seen from (\ref{K|2l}) rewritten in the present notation as
\begin{equation}
K_1\Bigr|_{2\ell}=\Pmatrix{
0 & \Po\Lo\Qo^T \\
\Qo\Lo\Po^T & 0
} ,
\label{eq:K|2loe}
\end{equation}
eq.\ (\ref{eq:formulaone}) multiplied by $\Po$ on the left,
$\Po\Lo\Qo^T\Qe=\Pe\Le$, just corresponds to the relation
$\K\vv_n=\kappa_n\uu_n$ following from (\ref{eq:Kf=kf}).\footnote{
Among the other three relations following from (\ref{eq:Kf=kf}),
the two corresponding to
$\K\uu_{n-\frac12}=\kappa_{n-\frac12}\vv_{n-\frac12}$ and the one with
$\uu$ and $\vv$ exchanged are trivial consequences of
(\ref{eq:O^TO=1}). However, the remaining one,
$\Qo\Lo\Po^T\Pe=\Qe\Le$ corresponding to $\K\uu_n=\kappa_n\vv_n$, does
not hold in the present regularization.
}

Let us mention another important formula concerning $\Pe$:
\begin{equation}
\Po^T\Pe\left(\Po^T\Pe\right)^T=1-\W ,
\label{eq:formulatwo}
\end{equation}
where the matrix $\W$ on the RHS is
\begin{equation}
\W_{nm}=\frac{2}{\pi^2}\frac{(-1)^{n+m}}{(n-1/2)(m-1/2)} .
\label{eq:W}
\end{equation}
Eq.\ (\ref{eq:formulatwo}) is easily proved from (\ref{eq:PoPe}).
As seen from the direct product form of $\W$ (\ref{eq:W}),
it is a projection operator of rank one:
\begin{equation}
\W^2=\W,\qquad \tr\W=1 .
\label{eq:W^2=W}
\end{equation}
Using $\Po\Po^T=1$, eq.\ (\ref{eq:formulatwo}) is rewritten into
\begin{equation}
\Pe\Pe^T=1-\Po\W\Po^T .
\label{eq:PePe^T}
\end{equation}

Having finished the derivation of the formulas of $P$ and $Q$,
let us turn to a recalculation of $G$.
Using the matrix representation (\ref{1/K}) and (\ref{eq:K|2loe})
and the fact that the $[1,1]$ component has contribution only from the
odd-odd sector, we can rewrite the first expression of (\ref{GofK})
for $G$ in terms of $P$ and $Q$ as
\begin{equation}
G=\frac{\pi}{4}\Bigl(\Po\Lo^{-1}\Po^T
-\Po\Lo\Qo^T\,\Qe\Le^{-3}\Qe^T\,\Qo\Lo\Po^T\Bigr)[1,1] .
\label{GPQ}
\end{equation}
Using (\ref{eq:formulaone}) and its transpose, eq.\ (\ref{GPQ}) can be
brought to an expression without $Q$:
\begin{align}
G&=\frac{\pi}{4}\Bigl(\Po\Lo^{-1}\Po^T -\Pe\Le^{-1}\Pe^T\Bigr)[1,1]
\nn\\
&=\frac{\pi}{\ln L}\sum_{n=1}^\infty\biggl(
\frac{1}{\kappa_{n-\frac12}}-\frac{1}{\kappa_n}\biggr) ,
\label{eq:Gagain}
\end{align}
where we have used that $\bigl(\overline{\uu}_n\bigr)_1
=\bigl(\overline{\uu}_{n-\frac12}\bigr)_1=2/\sqrt{\ln L}$.
Eq.\ (\ref{eq:Gagain}) is nothing but the previous (\ref{Gseries}).

\section{Energy density}
Now let us proceed to the evaluation of $H$.
Expressing $\Me$, $\Mo$ and $\vo$ in $H$ (\ref{eq:simpleH})
in terms of $\K$ and $\q$ by using
(\ref{M0simK1})--(\ref{eq:vo=Ku}) and keeping only those
terms with degree of divergence equal to three, we get
\begin{align}
H=\frac{\pi}{6}\,\q^T\K\Biggl\{&
\Kabsinv\biginv
-2\,\Kabsinv\K\Kabsinv\biginv\Kabsinv\K
\nn\\
&-3\,\K\Bigl(\Kabsinv\Bigr)^2\biginv\Kabsinv\K
+3\,\K\Bigl(\Kabsinv\Bigr)^3\K
\Biggr\}\K\q ,
\label{HKq}
\end{align}
with $\biginv$ for (\ref{HKq}) given by
\begin{equation}
\biginv=\biggl(1+\frac13\K\Kabsinv\K\Kabsinv\biggr)^{-1}.
\end{equation}
Using that the second term and the sum of the last two terms in
the curly bracket of (\ref{HKq}) are rewritten respectively into
\begin{equation}
-\Kabsinv\biginv\K\Bigl(\Kabsinv\Bigr)^2\K
-\biggl(\K\Bigl(\Kabsinv\Bigr)^2\K\Kabsinv\biginv\biggr)^T ,
\end{equation}
and
\begin{equation}
\K\Bigl(\Kabsinv\Bigr)^2\K\Kabsinv\biginv\K\Bigl(\Kabsinv\Bigr)^2\K ,
\end{equation}
we obtain a simpler expression of $H$:
\begin{align}
H=\frac{\pi}{6}\biggl(1-\K\Bigl(\Kabsinv\Bigr)^2\K\biggr)
\Kabsinv\biginv\biggl(1-\K\Bigl(\Kabsinv\Bigr)^2\K\biggr)[1,1].
\label{H11}
\end{align}

Now we use the matrix representations (\ref{1/K}) and
(\ref{eq:K|2loe}) for (\ref{H11}).
First, we have
\begin{align}
\biggl(1-\K\Bigl(\Kabsinv\Bigr)^2\K\biggr)_{oo}
&=1-\Po\Lo\Qo^T\,\Qe\Le^{-2}\Qe^T\,\Qo\Lo\Po^T
\nn\\
&=1-\Pe\Pe^T
\nn\\
&=\Po\W\Po^T ,
\end{align}
where we have used (\ref{eq:formulaone}) at the second equality,
and (\ref{eq:PePe^T}) in obtaining the last expression.
Next, using (\ref{eq:formulaone}) we have
\begin{align}
\biggl(\Kabsinv\biginv\biggr)_{oo}
&=\Po\Lo^{-1}\Po^T\left(
1+\frac13\,\Po\Lo\Qo^T\,\Qe\Le^{-1}\Qe^T\,\Qo\Lo\Po^T\,
\Po\Lo^{-1}\Po^T\right)^{-1}
\nn\\
&=\Po\Lo^{-1}\Po^T\left(
1+\frac13\,\Pe\Qe^T\Qo\Po^T\right)^{-1} .
\end{align}
Therefore, $H$ of (\ref{H11}) is rewritten into
\begin{equation}
H=\frac{\pi}{6}\Po\W\Lo^{-1}\left(
1+\frac13\,\Po^T\Pe\,\Qe^T\Qo\right)^{-1}\W\Po^T
\,[1,1] .
\label{eq:H111}
\end{equation}
Then, using (\ref{eq:PoPe}), (\ref{eq:QoQe}) and
\begin{equation}
\left(\W\Po^T\right)_{n1}
=\sum_{m=1}^\infty\W_{nm}\frac{2}{\sqrt{\ln L}}
=-\frac{2}{\sqrt{\ln L}}\,\frac{(-1)^n}{\pi(n-1/2)} ,
\end{equation}
we obtain the final expression of $H$:
\begin{equation}
H=\frac{1}{3\pi^2}\sum_{n,m=1}^\infty
\frac{1}{(n-1/2)^2}\left(A^{-1}\right)_{nm}
\frac{1}{m-1/2} ,
\label{Hfinal}
\end{equation}
where the matrix $A$ is given by
\begin{equation}
A_{nm}=\delta_{n,m}+\frac{4}{3\pi^2}\left(n-\frac12\right)
\sum_{k=1}^\infty\frac{k}{\left[(n-1/2)^2-k^2\right]
\left[(m-1/2)^2-k^2\right]} .
\label{eq:A}
\end{equation}

Now we have obtained a largely simplified expression of $H$ compared
with the original (\ref{eq:simpleH}) or (\ref{HKq}).
The main difference between the original expression (\ref{HKq})
for $H$ and the present one (\ref{Hfinal}) is that,
although each term in (\ref{HKq}) does contain divergence and they
cancel as a whole, our final expression (\ref{Hfinal})
is a well-defined infinite series without containing any divergences.
Unfortunately, we have not succeeded in evaluating the infinite series
analytically. Instead, we have carried out numerical calculation of
(\ref{Hfinal}) by reintroducing the cutoff $L$ to the infinite
summations in (\ref{Hfinal}) and (\ref{eq:A}).
The result given in table \ref{tab:H} suggests very strongly
that $H=(1/2)\ln(27/16)$, in agreement with the result of \cite{BCFT}.
Therefore, the interpretation of the present classical solution
$\Psic$ as the configuration of two D$25$-branes \cite{HatMor} is
rejected.
\begin{table}[htbp]
\begin{center}
\parbox{5cm}{
\begin{tabular}[b]{|r|c|}
\hline
$L$~ & $H/[\ln(27/16)/2]$\\
\hline\hline
$50$ & $0.9999481903$ \\
$100$ & $0.9999869824$ \\
$150$ & $0.9999942047$ \\
$200$ & $0.9999967374$ \\
$250$ & $0.9999979109$ \\
$300$ & $0.9999985488$ \\
\hline
\end{tabular}
}
\caption{Numerical values of $H$ (\ref{Hfinal}) for various cutoff $L$.
}
\label{tab:H}
\end{center}
\end{table}

\section{Conclusion}
In this paper we have shown how the twist anomaly is evaluated
analytically.
During our analysis, we found how the twist anomaly is
realized in the finite-size matrix regularization.
Naively, the eigenvalues of the Neumann coefficient matrix $\Me$
degenerate due to the twist symmetry.
However, after introducing the regularization, the twist symmetry
breaks down and the degeneracy of the eigenvalues is lifted.
The quantity $G$ was evaluated exactly to reproduce the expected tachyon
mass squared of $-1$.
On the other hand, for the quantity $H$ related to the energy density,
we obtained a simple expression as a well-defined infinite series,
though we could not evaluate its value analytically.
We shall conclude our paper by presenting several further directions
of our analysis.
\begin{itemize}
\item
As a technical problem, we have to evaluate the infinite series
(\ref{Hfinal}) for $H$ analytically.
Proof of the eigenvalue spectrum (\ref{Lambda2l}) of
$\K\bigl|_{2\ell}$, which is an assumption in this paper, is also
a remaining subject.
It is also necessary to give a rigorous justification to the
prescription of keeping only terms with degree of divergence equal to
three (see sec.\ 2.2).

\item
In the usual terminology, anomaly appears when there are no
regularizations compatible with all the symmetries.
We would like to understand the twist anomaly in the same sense.
Especially, it should be important to understand which symmetries our
finite-size matrix regularization respects.

\item
We evaluated the twist anomaly by adopting the regularization of
truncating the size of the infinite matrices.
We would also like to derive the same results as obtained in this
paper by a more refined and more systematic regularization.
Methods of \cite{Oku3} would be an interesting possibility.

\item
In \cite{BCFT} and our present analysis, it was found that the energy
density of the solution $\Psic$ does not reproduce the correct
D$25$-brane tension.
In \cite{BCFT} it was further pointed out that the reason why the
energy density deviates from the expected value is that the present
tachyon wave function $\Phit$ does not satisfy the linearized equation
of motion $\calQB\Phit=0$ (\ref{eq:QBPhit}) in the strong sense.
Having seen the correspondence of the final results between our
algebraic analysis \cite{HatKaw,HatMor} and the geometrical approach
\cite{BCFT}, it is an urgent task to obtain the tachyon wave function
which satisfy the linearized equation of motion in the strong sense
and at the same time reproduces the correct D$25$-brane tension.
A root of the problem in the geometric approach lies in the fact that
the cubic product $\Phi_1\cdot(\Phi_2 *\Phi_3)$ of three sliver-based
states $\Phi_k$ with momentum insertion depends on how we take the
limit $n_k\to\infty$ \cite{BCFT,RasVis}, where we express the state
$\Phi_k$ as $n_k$-wedge state \cite{sliver}.

\item
As far as we have investigated, physical observables in VSFT
are always related to quantities that naively vanish.
We would like to understand the deep reason of this phenomenon.
It might be related to the fact that in VSFT (\ref{eq:SV}) expanded
around $\Phi=0$ there are no physical excitations since the kinetic
term consists only of the purely ghost operator.

\item
Note that the twist anomaly are written in terms of the Neumann
coefficients, which express the open string interactions.
This indicates that the twist anomaly might be a fundamental
phenomenon which appears universally in the open string interactions.
Usually the scattering amplitudes in CSFT have been calculated by
mapping them to a complex plane. However, recalling that the twist
anomaly can also be evaluated on the complex plane \cite{BCFT}, the
string amplitudes calculated thoroughly in terms of the Neumann
coefficients might be recognized as twist anomaly (though this
expectation contradicts the one mentioned in the above item).
We would also like to know how the recent explanation of the
emergence of closed strings in VSFT \cite{GRSZ} is related to twist
anomaly.

\end{itemize}

\section*{Acknowledgments}
We would like to thank I.\ Kishimoto, H.\ Kogetsu and H.\ Ooguri for
valuable discussions and comments.
The works of H.\,H.\ and S.\,M.\ were supported in part by a
Grant-in-Aid for Scientific Research from Ministry of Education,
Culture, Sports, Science, and Technology (\#12640264 and \#04633,
respectively).
S.\,M.\ is supported in part by the Japan Society for
the Promotion of Science under the Predoctoral Research Program.

\appendix
\section{Proof of the vector formulas (\ref{v0}) and (\ref{v1})}
\label{app:vector}

In this appendix, we present a proof of the formulas (\ref{v0}) and
(\ref{v1}) which express $\ve$ and $\vo$ in terms of a simpler vector
$\q$. Since (\ref{v0}) is not directly used in this paper and
its proof is almost the same as that for (\ref{v1}),
we shall mainly concentrate on proof of the formula (\ref{v1}).
Our argument is similar to that for the eigenvector of $\Me$
corresponding to the eigenvalue $-1/3$ given in sec.\ 3 of
\cite{spec}.

The original matter Neumann coefficients have the following
integral representation \cite{LPP1,LPP2}:
\begin{align}
&V^{rs}_{nm}=-\frac{1}{\sqrt{nm}}\resint{z}\resint{w}
\frac{1}{z^n w^m}\frac{f'_r(z)f'_s(w)}{\left(f_r(z)-f_s(w)\right)^2} ,
\label{eq:Vnm}
\\
&V^{rs}_{n0}=-\frac{1}{\sqrt{n}}\resint{z}
\frac{1}{z^n}\frac{f'_r(z)}{f_r(z)-f_s(0)} ,
\label{eq:Vn0}
\end{align}
where $f_r(z)$ is given by\footnote{Note that, compared
with the formulas in \cite{LPP1,LPP2}, $\omega$ is replaced with
$1/\omega$.}
\begin{equation}
f_k(z)=f(z)\,\omega^{-k} ,
\label{eq:f_k}
\end{equation}
with
\begin{equation}
f(z)=\left(\frac{1+ iz}{1-iz}\right)^{2/3},
\quad
\omega=e^{2\pi i/3} .
\end{equation}
The integration contours in (\ref{eq:Vnm}) and (\ref{eq:Vn0}) are
circles around the origin.
These Neumann coefficient matrices are related to our present ones
by\footnote{
The Neumann coefficient $V_{n0}^{rs}$ is unique only when the index
$s$ is contracted with a conserved quantity $\alpha_s$ satisfying
$\sum_{s=1}^3\alpha_s=0$.
The vector $\ve$ in this paper in a generic representation of
$V_{n0}^{rs}$ is
$-(1/3)(\bm{v}_{+0}+\bm{v}_{-0})$ in \cite{HatKaw}, namely, the one
defined by (\ref{eq:v=Vn0}).
This is a representation independent quantity.
Only when $V_{n0}^{rs}$ is defined through the $6$-string Neumann
coefficient \cite{GroJev1,GroJev2}, we have $(\ve)_n=V_{n0}^{rr}$
since (\ref{eq:v+v+v=0}), i.e.\
$(V^{rr}+V^{r,r+1}+V^{r,r-1})_{n0}=0$, holds in this representation.
In the text we are taking the representation in terms of the
$6$-string Neumann coefficient.
}
\begin{align}
&(C\Me)_{nm}=V^{rr}_{nm},
\qquad(C\Mo)_{nm}=(V^{r,r+1}-V^{r,r-1})_{nm},\\
&(\ve)_n=\frac13\left(2 V^{rr}-V^{r,r+1}-V^{r,r-1}\right)_{n0},
\quad(\vo)_n=(V^{r,r+1}-V^{r,r-1})_{n0}.
\label{eq:v=Vn0}
\end{align}
Especially, the integral representation of $C\Mo$ and $\vo$ are given
as
\begin{align}
(C\Mo)_{nm}&=-\frac{4 i}{3}\sqrt{\frac{m}{n}}\resint{z}\resint{w}
\frac{f(z)}{z^n w^{m+1}(1+z^2)}
\left[\frac{1}{f(z)-\omega^* f(w)}
-\frac{1}{f(z)-\omega f(w)}\right] ,
\label{CM1}\\
(\vo)_n&=-\frac{4 i}{3}\frac{1}{\sqrt{n}}\resint{z}
\frac{f(z)}{z^n(1+z^2)}\left[
\frac{1}{f(z)-\omega^*}-\frac{1}{f(z)-\omega}
\right] .
\label{eq:v1}
\end{align}
In deriving (\ref{CM1}) we have carried out an integration by parts
with respect to $w$.
To prove the formula (\ref{v1}), let us calculate $M_1\q$.
Since we have $\Mo\q=-C\Mo\q$ owing to the twist property $C\Mo
C=-\Mo$ and $C\q=\q$, we shall calculate $-C\Mo\q$.

To make the following calculation well-defined, we use the regularized
version of $\q$ instead of the original one (\ref{eq:u}):
\begin{equation}
q_n=\frac{a^{-n-1}}{2\sqrt{n}}\left[i^n+(-i)^n\right],
\quad(a\to 1+0) .
\label{eq:regq}
\end{equation}
Then, from (\ref{CM1}), (\ref{eq:regq}) and the geometric series
\begin{equation}
\sum_{m=1}^\infty\frac{i^m+(-i)^m}{2\,(aw)^{m+1}}
=-\frac{1}{aw(1+a^2w^2)},
\label{eq:GS}
\end{equation}
we obtain
\begin{equation}
\left(M_1\q\right)_n
=-\frac{4i}{3}\frac{1}{\sqrt{n}}\resint{z}
\frac{f(z)}{z^n(1+z^2)}F_1(z) ,
\label{eq:M1q}
\end{equation}
with $F_1(z)$ defined by
\begin{equation}
F_1(z)=\resint{w}\frac{1}{aw(1+a^2w^2)}
\left[\frac{1}{f(z)-\omega^* f(w)}
-\frac{1}{f(z)-\omega f(w)}\right] .
\label{eq:F1}
\end{equation}
The integration contour for (\ref{eq:F1}) must satisfy $|w|>1/a$ due
to the convergence requirement of the series (\ref{eq:GS}).
The integration $F_1(z)$ has contributions from poles at
$w=0$ and $\pm i/a$ (the pole at $w=-1/z$ corresponding to
$f(z)-\omega^{\pm}f(w)=0$ is outside the $w$-integration contour),
and we have
\begin{align}
aF_1(z)
&=\frac{1}{f(z)-\omega^*}-\frac{1}{f(z)-\omega}
\nn\\
&\qquad+\sum_\pm
\frac{1}{(\pm i/a)2a^2(\pm i/a)}\left[
\frac{1}{f(z)-\omega^* f(\pm i/a)}
-\frac{1}{f(z)-\omega f(\pm i/a)}\right]
\nn\\
&\underset{a\to 1}{\longrightarrow}
\frac{1}{f(z)-\omega^*}-\frac{1}{f(z)-\omega} .
\label{eq:aF1}
\end{align}
Note that both of the terms coming from $w=\pm i/a$ vanish in the
limit
$a\to 1$.

Comparing (\ref{eq:M1q}) for $M_1\q$ with $F_1(z)$ given by
(\ref{eq:aF1}) and the integral representation (\ref{eq:v1}) of $\vo$,
we find that they are equal:
\begin{align}
\vo=M_1\q .
\label{eq:v1=M1q}
\end{align}
We can prove the formula (\ref{v0}) in quite a similar way.

\section{Inner product of the eigenvectors}
\label{app:norm}

In this appendix we calculate the inner product of the eigenvectors of
the matrix $\K$.\footnote{A similar derivation of the inner products has
been given in \cite{Oku2}.
However, since we need the inner product of finite-$L$ truncated
eigenvectors, we shall rederive the inner product in a form applicable
to the calculations in the text.
}
The eigenvalue problem of the matrix $\K$ has been solved in
\cite{spec}.
There, the matrix $\K$ is represented as a differential operator
$-(1+z^2)(d/dz)$ acting on the function $f(z)$ made from a generic
vector $\bm{f}=(f_n)$:
\begin{equation}
f(z)=\sum_{n=1}^\infty\frac{f_n}{\sqrt{n}}z^n,
\label{eq:fpseries}
\end{equation}
and the eigenvalues and the eigenvectors are obtained by solving
differential equations.
The eigenvalues of the matrix $\K$ range over the real axis uniformly.
The function $f^{(\kappa)}(z)$ corresponding to the eigenvector
$\bm{f}^{(\kappa)}$ with eigenvalue $\kappa$ is given by
\begin{equation}
f^{(\kappa)}(z)
=\frac{1}{\kappa}\Bigl(1-\exp(-\kappa\tan^{-1}z)\Bigr)
=z-\frac{\kappa}{\sqrt{2}}\frac{z^2}{\sqrt{2}}
+\frac{\kappa^2-2}{2\sqrt{3}}\frac{z^3}{\sqrt{3}}+\cdots.
\label{eq:fz}
\end{equation}
{}From (\ref{eq:fz}) the eigenvectors $\bm{f}^{(\kappa)}$ before
normalization can be read off as
\begin{equation}
\bm{f}^{(\kappa)}
=\Bigl(1,-\frac{\kappa}{\sqrt{2}},\frac{\kappa^2-2}{2\sqrt{3}},
\cdots\Bigr),
\label{fcomp}
\end{equation}

\begin{figure}[htbp]
\begin{center}
\leavevmode
\epsfxsize=4cm
\epsfbox{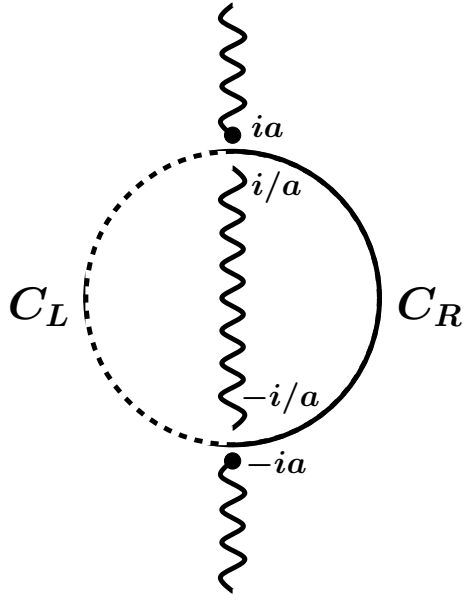}
\put(-50,174){$\bm{ia}$}
\put(-50,152){$\bm{i/a}$}
\put(-54,72){$\bm{-i/a}$}
\put(-52,47){$\bm{-ia}$}
\put(5,105){{\Large$\bm{C_R}$}}
\put(-142,105){{\Large$\bm{C_L}$}}
\caption{The contour of the integration (\ref{eq:regint}) to evaluate
the norm of $\bm{f}$.
The blobs at $z=\pm ia$ are the poles of the integrand of
(\ref{eq:regint}).
}
\label{figone}
\end{center}
\end{figure}

The inner product between two generic vectors $\bm{f}$ and $\bm{g}$ is
defined by
\begin{equation}
\bm{f}\cdot\bm{g}\equiv\sum_{n=1}^\infty f_n g_n.
\end{equation}
It is expressed by a contour integral using the corresponding
functions $f(z)$ and $g(z)$ as
\begin{equation}
\bm{f}\cdot\bm{g}=\oint_{|z|=1}
\frac{dz}{2\pi i}\biggl(\frac{d}{dz}f(z)\biggr)g(1/z).
\end{equation}
In particular, the inner product between the eigenvectors
is given by
\begin{equation}
\bm{f}^{(\kappa)}\cdot\bm{f}^{(\lambda)}=
\frac{-1}{\lambda}\oint_{|z|=1}
\frac{{\rm d}z}{2\pi i}\frac{1}{1+z^2}
\exp\Bigl(-\kappa\tan^{-1}z\Bigr)
\exp\Bigl(-\lambda\tan^{-1}\frac1z\Bigr).
\label{formalintegral}
\end{equation}
However, this integral is not well-defined since there exist poles
at $z=\pm i$ and some branch-cuts.
In order to treat $\tan^{-1}z=(1/2i)\ln\left[(1+iz)/(1-iz)\right]$
properly, let us take the branch-cut of $\ln z$ to be $\Im z=0$, $\Re
z<0$.
Then the branch-cuts of $\tan^{-1}z$ and $\tan^{-1}{1/z}$ runs
over the imaginary axis and these cuts meet at $z=\pm i$.
Therefore,
we adopt the same regularization as used in (\ref{eq:regq}).
Namely, we deform the function $f^{(\kappa)}(z)$ into
\begin{equation}
f^{(\kappa)}(z)=\frac{1}{\kappa}
\biggl(1-\exp\Bigl(-\kappa\tan^{-1}\frac{z}{a}\Bigr)\biggr)
=\sum_{n=1}^\infty\frac{f_n}{\sqrt{n}}\biggl(\frac{z}{a}\biggr)^n,
\end{equation}
with $a=1+0$.
This deformation corresponds to replacing $f_n$ with $f_n/a^n$,
which serves effectively as truncation of the infinite dimensional
vector into a finite $L$-dimensional one with
\begin{equation}
a^L\simeq e .
\label{finiteL}
\end{equation}
On the other hand, this deformation slightly moves the poles and the
endpoints of the branch-cuts as in fig.\ \ref{figone}.
Since the branch-cuts of $\tan^{-1}z$ ($\tan^{-1}{1/z}$)
runs from $z=ia$ ($i/a$) to $z=-ia$ ($-i/a$) along the imaginary
axis, the integral along the contour $|z|=1$ is quite safe:
\begin{equation}
\bm{f}^{(\kappa)}\cdot\bm{f}^{(\lambda)}=
\frac{-1}{\lambda}\oint_{|z|=1}
\frac{dz}{2\pi i}\frac{a}{a^2+z^2}
\exp\biggl(-\kappa\tan^{-1}\frac{z}{a}\biggr)
\exp\biggl(-\lambda\tan^{-1}\frac{1}{az}\biggr) .
\label{eq:regint}
\end{equation}
Next, we separate the contour of (\ref{eq:regint}) into two
segments $C_R$ and $C_L$,
with $C_{R(L)}$ being the parts of the original contour $|z|=1$ on the
right (left) half plane.
Using the identity
\begin{equation}
\tan^{-1}z+\tan^{-1}\frac{1}{z}=
\begin{cases}
\pi/2 & \Re z>0 \\
-\pi/2 & \Re z<0
\end{cases},
\label{arctan id}
\end{equation}
we have
\begin{align}
\bm{f}^{(\kappa)}\cdot\bm{f}^{(\lambda)}=&\frac{-1}{\lambda}\biggl[
\exp\biggl(-\frac{\lambda\pi}{2}\biggr)\int_{C_R}\frac{dz}{2\pi i}
+\exp\biggl(\frac{\lambda\pi}{2}\biggr)\int_{C_L}\frac{dz}{2\pi i}
\biggr]\frac{a}{a^2+z^2}
\exp\biggl(-(\kappa-\lambda) \tan^{-1}\frac{z}{a}\biggr)\nn\\
&\qquad\times\exp\biggl(-\lambda\Bigl(\tan^{-1}\frac{z}{a}
-\tan^{-1}az\Bigr)\biggr).
\label{eq:fcdotf}
\end{align}
The difference
\begin{equation}
\tan^{-1}\frac{z}{a}-\tan^{-1}az ,
\label{difference}
\end{equation}
vanishes\footnote{Strictly speaking, this difference has a non-zero
value near $z=\pm i$. However, we can show that it does not
contribute to the total integral.}
in the limit of $a\rightarrow1$.
Therefore, dropping the final exponential factor in (\ref{eq:fcdotf}),
we obtain
\begin{align}
\bm{f}^{(\kappa)}\cdot\bm{f}^{(\lambda)}
&=\frac{2\sinh(\lambda\pi/2)}{\lambda\pi}
\frac{1}{\kappa-\lambda}
\sin\biggl(\frac{\ln L}{2}(\kappa-\lambda)\biggr) ,
\label{inner}
\end{align}
where we have used the relation (\ref{finiteL}) between $a$ and $L$.
If we further use one of definitions of the delta function:
\begin{align}
\pi\delta(\kappa)=\lim_{L\to\infty}
\frac{1}{\kappa}\sin{\frac{\ln L}{2}\kappa},
\label{delta}
\end{align}
we find that in the limit $L\to\infty$
\begin{align}
\bm{f}^{(\kappa)}\cdot\bm{f}^{(\lambda)}=
\frac{2\sinh(\lambda\pi/2)}{\lambda}\,
\delta(\kappa-\lambda).
\label{norm}
\end{align}

\end{document}